\documentclass{aa}
\usepackage{graphicx}
\usepackage{natbib}
\bibpunct{(}{)}{;}{a}{}{,}
\usepackage{txfonts}
\begin{document}
   \title{Multi-wavelength analysis of the dust emission \\ in the Small
     Magellanic Cloud\thanks{based on observations with ISO, an ESA project with instruments funded by ESA Member States (especially the PI countries: France, Germany, The Netherlands and the UK) and with the participation of ISAS and NASA.}}
   \author{C.~Bot
          \inst{1}\fnmsep\inst{2}
          \and
          F.~Boulanger\inst{2}
	  \and
          G.~Lagache\inst{2}
          \and
	  L.~Cambr\'esy\inst{1}
          \and
          D.~Egret\inst{3}
          }

   \offprints{C.Bot,
   \email{bot@astro.u-strasbg.fr}}

   \institute{UMR 7550, Centre de Donn\'ees Astronomiques de Strasbourg (CDS), Universit\'e Louis Pasteur, F-67000 STRASBOURG, France
         \and
             Institut d'Astrophysique Spatiale, Universit\'e
             Paris-Sud, F-91405 Orsay, France
          \and
             UMS 2201, Observatoire de Paris, F-75014 Paris, France
             }

   \date{Received ...; accepted ...}

   \abstract{We present an analysis of dust grain emission in the diffuse interstellar medium of the Small Magellanic Cloud (SMC). 
This study is motivated by the availability of 170~$\mathrm{\mu}$m ISOPHOT data covering a large part of the SMC, with a resolution enabling to disentangle the diffuse medium from the star forming regions. 
After data reduction and subtraction of Galactic foreground emission, we used the ISOPHOT data together with HiRes IRAS data and ATCA/Parkes combined \ion{H}{i} column density maps to determine dust properties for the diffuse medium. 
We found a far infrared emissivity per hydrogen atom 30 times lower than the Solar Neighborhood value. The modeling of the spectral energy distribution of the dust, taking into account the enhanced interstellar radiation field, gives a similar conclusion for the smallest grains (PAHs and very small grains) emitting at shorter wavelength. Assuming Galactic dust composition in the SMC, this result implies a difference in the gas-to-dust ratio (GDR) 3 times larger than the difference in metallicity.
This low depletion of heavy elements in dust could be specific of the diffuse ISM and not apply for the whole SMC dust if it results from efficient destruction of dust by supernovae explosions.
  \keywords{ dust -- Magellanic Clouds -- ISM: abundances -- evolution -- Galaxies: ISM }
   }

   \titlerunning{Multi-wavelength analysis of the dust emission in the SMC}
   \authorrunning{C.Bot et al.}
   \maketitle
%

\section{Introduction}

The Small Magellanic Cloud (SMC) is a dwarf irregular galaxy at a distance of approximately 60~kpc \citep{Wes91}. 
Its relative proximity enables to have a global view of the galaxy, as well as a sufficiently good resolution to disentangle the diffuse medium from the star forming regions and molecular clouds. The SMC is an ideal target to study the low metallicity diffuse medium present not only in dwarf galaxies, but also in the external parts of large spirals.

The SMC contains several \ion{H}{ii} regions and molecular clouds \citep{RLB93}\nocite{RLB+93}. It has a high interstellar radiation field (ISRF), between 4 and 10 times that of the Solar neighborhood \citep{Leq79,VLMR80,ALM88}. 
It is extremely rich in neutral atomic gas \citep{SKPS98} and has a low heavy element abundance \citep{Duf84,SV91}. All of these characteristics are indicative of a `` young'' galaxy, actively forming stars, which might in some sense be considered as a local model of distant galaxies at the beginning of their evolution.

Dust grains are composed mainly of heavy atoms (e.g. carbon, silicium, oxygen, \ldots). 
In a low metallicity environment, such elements are lacking. 
How does this deficiency act on the dust grain abundances? 
 The gas-to-dust ratio (GDR) is often assumed to be proportional to the metallicity (i.e. there is a fixed fraction of heavy elements in dust).
Based on a model of the evolution of elemental abundances in gas and dust, \citet{Dwe98} concludes that this proportionality applies across the disk of the Galaxy. 
However, observations in dwarf irregular and blue compact dwarf galaxies show a large scatter in the GDR versus metallicity relation. 
This indicates that a simple scaling might be a too simplistic description of the phenomenon \citep{IMW90,LF97,LF98}.
The GDR is expected to vary among ISM components because destruction of dust by shock waves is more efficient in the diffuse gas while accretion of heavy elements on dust dominates in molecular clouds.
The integrated GDR is thus expected to depend on the relative fraction of matter in diffuse/dense components and on the supernova rate and thereby on the star formation history \citep{Hir99}.

The goal of our study is to characterize the dust properties in the diffuse medium of the SMC, and in particular the GDR.
Comparing the GDR and the dust spectral energy distribution for the SMC and the Galaxy, we quantify the effect of the difference in  metallicity on dust abundance and size distribution (e.g. the relative amounts of PAHs, very small grains and big grains). 

Since the gas and the dust masses are closely mixed, where the radiation is homogeneous the dust emission is expected to correlate with the distribution of the gas. In the diffuse medium, gas is mainly composed of neutral hydrogen and the GDR can be derived from the comparison between the dust infrared emission with the \ion{H}{i} column density, provided that the dust temperature is known. This is not the case in star-forming regions, where the infrared emission from dust in \ion{H}{ii} regions and molecular clouds closely associated with luminous stars can become dominant.
Several authors have computed GDRs, comparing the IRAS infrared emission with the \ion{H}{i} column density, taking the $100 \mathrm{\mu m}/60 \mathrm{\mu m}$ intensity ratio to derive the dust temperature \citep{SI89,SVT90,SSV+00}, but this ignores the contribution from small grains at $60\mathrm{\mu m}$. An additional difficulty comes from the variations in the intensity of the radiation field inside a beam, which casts doubt on any GDR estimate from studies where the angular resolution is too coarse to separate star forming regions from the diffuse emission \citep{LD02,ABC+03}

The availability of 170~$\mathrm{\mu}$m ISOPHOT observations enables us to have a new view on dust properties of the SMC. 
Independently of our work, \citet{WSH+03} have also reduced and analyzed these data but with a distinct focus on point sources. In a second paper \citep{WKL+04}, these authors adress the same question as us but they investigate the main body (bar) of the SMC without discarding the star forming regions.

After data reduction and validation (Sect. \ref{sec:data} and \ref{sec:phot_cons}), the 170~$\mathrm{\mu}$m map is compared to additional observations (Sect. \ref{sec:projres}), in particular the IRAS High Resolution map (HiRes) at 100~$\mathrm{\mu}$m. 
Having obtained the 170/100~$\mathrm{\mu}$m emission ratio, in Sect. \ref{sec:IRcolors} we derive a reference temperature for the big grains in the diffuse medium of the SMC.
An \ion{H}{i} column density map from ATCA/Parkes combined data \citep{SSD+99} is used to study the dust to gas correlation, allowing us to estimate the emissivity of the dust per hydrogen atom for the diffuse medium (Sect.\ref{sec:gas-dust-col}) . 
Finally, the spectral energy distribution of the dust grains in the diffuse medium of the SMC is fitted with the \citet{DBP90} model to quantify the abundances of the different grain components (Sect. \ref{sec:diffhi}).
The results are discussed in Sect. \ref{sec:discuss}; Sect. \ref{sec:ccl} presents our conclusions.

\section{The Data}\label{sec:data}

This study relies on archive data. 
Some were ready-made as the IRAS maps at 12, 25, 60, and 100~$\mathrm{\mu m}$, or the ATCA/Parkes combined data of the \ion{H}{i} line at 21~cm.
 The ISOPHOT data was imported from the ISO Data Archive\footnote{http://iso.vilspa.esa.es/ida/index.html}.
 But the quality of the ISOPHOT map at 170~$\mathrm{\mu}$m in the ISO archive was not sufficient for this study, so we retrieved, reduced and calibrated the data.
 We first present the ISOPHOT data and the reduction processes. Then we describe the ancillary data that were used.

\subsection{The ISOPHOT data}

The Small Magellanic Cloud was almost fully covered by 10 observations with the ISOPHOT instrument \citep{LKA+96} on board the ISO satellite \citep{KSA+96,LKR+03}.
 The observations were made using the P22 AOT (Astronomical Observation Template), i.e. a raster map on a two-dimensional regular grid, with the C200 detector (2x2 pixels) at an effective wavelength of $170~\mathrm{\mu}$m.
 The pixel field of view is 1.5~$\arcmin$ and the raster pointing spacings are 180$\arcsec$ in both in-scan and cross-scan directions.
 The ISOPHOT point spread function can be approximated by a 2D gaussian with 90~$\arcsec$ full width half maximum (FWHM) in both directions and is supposed to be constant across the map. 

We reduced the data with PIA\footnote{The ISOPHOT data presented in this paper were reduced using PIA, which is a joint development by the ESA Astrophysics Division and the ISOPHOT Consortium with the collaboration of the Infrared Processing and Analysis Center (IPAC).
 Contributing ISOPHOT Consortium institutes are DIAS, RAL, AIP, MPIK, and MPIA.} (ISOPHOT Interactive Analysis) V10.0 \citep{GAK+97,GAH+98}, except for the flat-field correction. 
In the following, we describe the specific treatments applied.
Some observations showed specific problems on raw data (e.g. a pixel went into erratic state or a severe glitch affected a pixel until the end of the observation).
We choose to remove the data for these pixels in the corresponding observations.
If this removal is not done, the reduction processes will interpret these erratic fluxes as flat-fielding. This distorts the photometry of the other pixels for the whole observation.

To deal with the discarded pixels, we choose to perform the flat-field correction without the PIA tool as in \citet{LD01}.
The flat-field correction is on average $1.01\pm 0.05$, $0.93\pm0.03$, $1.17\pm0.09$ and $0.92\pm0.03$ for pixel 1, 2, 3 and 4 respectively.
Reproducible flat-field correction values for the ISOPHOT C\_200 array have been published by \citet{LD01} for the carefull reduction of 3 deep fields (FIRBACK).
 We do not observe the same average flat field corrections factor and we find a higher dispersion from one observation to another.
 This difference may be because the responsivity of ISOPHOT C\_200 pixels changes with the observed intensity, so that the flat-field correction computed with low fluxes is no longer valid for the brightest pixels \citep{PLB+03}.
 This effect might affect the 170~$\mathrm{\mu}$m flux determination for bright sources.
 The reproducible behavior of pixels observed by \citet{LD01} is then due to the relative flatness of deep fields.

The different observations have been done at different times, therefore the ``absolute calibration'' performed using the FCS measurement can be slightly different from one raster to another.
 This difference is observed as discontinuities between maps in the mosaic.
 To correct for these small absolute calibration differences (for a border between two maps, the mean difference in brightness observed is about 13\%), we applied a `` general flat-field'' correction as in \citet{LD01}.

\begin{figure*}
	\centering
		\includegraphics[angle=90,width=17cm]{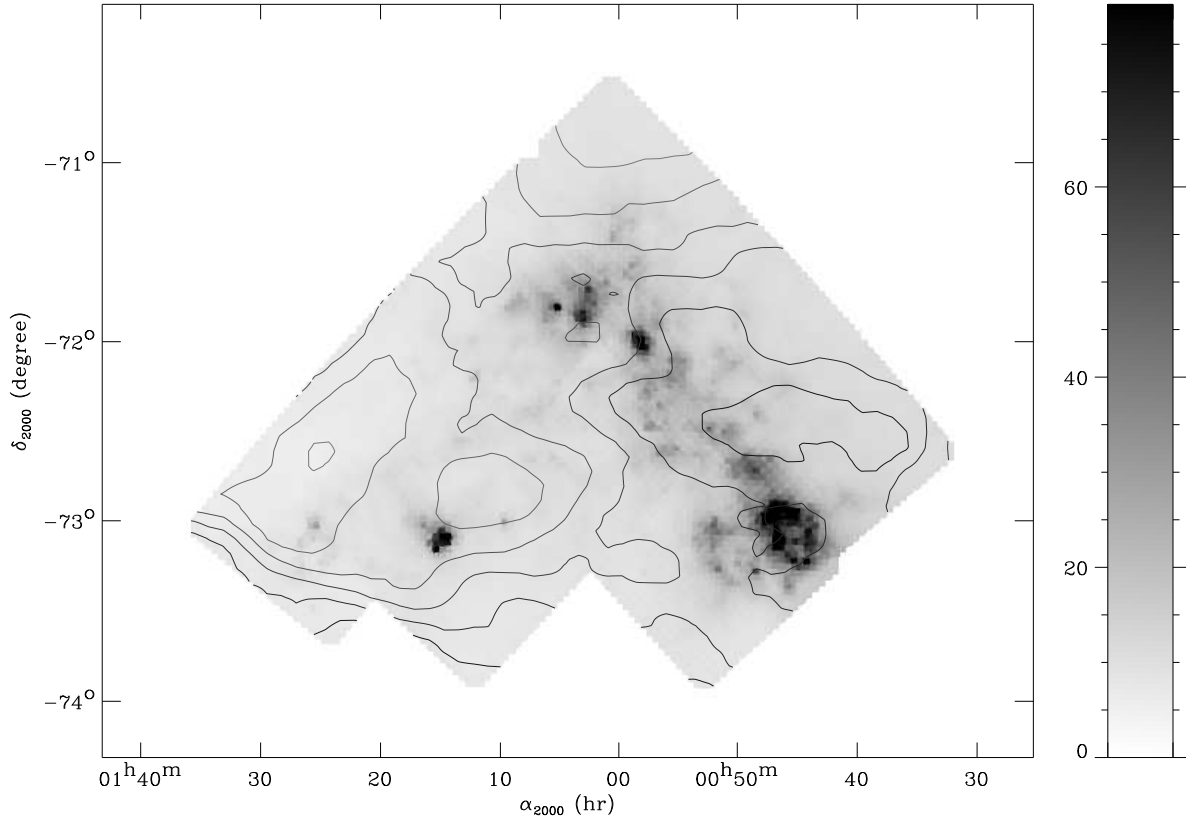}
	\caption{The ISOPHOT map at 170~$\mathrm{\mu}$m (in MJy/sr) after reduction processes and a bilinear interpolation to fill the holes left by the mapping pattern. The flat-field has been efficiently corrected for the diffuse part of the map, but one can still see low level defects in the brightest parts. Contours of the \ion{H}{i} Galactic foreground are overplotted at column densities of 2.6, 2.8, 3., 3.2, 3.4 and 3.6$\cdot 10^{20}$ H atom.cm-2 (from light grey to black). \label{fig:map_pht}}
\end{figure*}

After concatenating the timelines, the raster is projected on the sky using each individual pixel coordinates, with a pixel size for the map of 10~$\arcsec$ to match the IRAS HiRes gridding.
 Because the raster step size of each observation was a full detector size, and the spacecraft Y-axis was not parallel to right ascensions or declinations, we observe gaps between the individual array pointings.
 The final map is presented at Fig. \ref{fig:map_pht} after a bilinear interpolation to reconstruct the signal in the gaps. 

\subsection{Ancillary data}

\subsubsection{IRAS HiRes data}\label{sec:hires}

As a complement to the ISOPHOT data at 170~$\mathrm{\mu}$m, we used IRAS high resolution (HiRes) data at 12, 25, 60 and 100 $\mathrm{\mu}$m. 
Maps of $2\degr\times2\degr$ centered at $00^h 57^m 29.12^s$  $-72\degr 32\arcmin 29.2\arcsec$ in J2000 were obtained. 
The pixel size is 10~$\arcsec$. 
The HiRes point spread function (PSF) varies spatially in a map. 
A gaussian approximation of the PSF is given with the HiRes maps and is defined by two spatially dependent FWHMs and a rotation angle. 
For more information on HiRes, please refer to \citet{AFM90} and \citet{BKK94}.
Since we are searching for extended structure properties, the HiRes IRAS maps were brought to the DC-calibration \citep{WGC+94}.
This DC calibration is only valid for spatial scales greater than 40~\arcmin.
 As the SMC exhibits structure on various spatial scales, the DC calibration introduces some limitations that will be discussed in Sect. \ref{sec:phot_iras}.

\subsubsection{ATCA/Parkes combined SMC \ion{H}{i} data}

To study the correlation between gas and dust in the SMC, we needed a \ion{H}{i} column density map to compare with the ISOPHOT map, in order to compute the abundances of dust grains relative to the hydrogen.

\citet{SSD+99} have combined Parkes telescope observations of neutral hydrogen in the SMC with the Australia Telescope Compact Array (ATCA) aperture synthesis mosaic to obtain a set of images sensitive to all angular spatial scales between 98~$\arcsec$ and $4\degr$ from 90 to 215~km.s$^{-1}$. The spatial resolution obtained is 98~$\arcsec$ and the velocity resolution is 1.65~km.s$^{-1}$. We used a column density map of the SMC, obtained by integrating over the whole velocity range and assuming that the neutral hydrogen is optically thin\footnote{The column density map and the data cube are available at \rm{ftp://ftp2.naic.edu/pub/ast/sstanimi/.}}.

\subsubsection{Parkes \ion{H}{i} data for the Galactic foreground}\label{sec:data_gal_fil}

A Galactic filament exists in front of the SMC. Even though it is very smooth, large (see contours of Fig. \ref{fig:map_pht}) and uniform (column densities between $2.5\cdot 10^{20}$ and $4\cdot 10^{20}$ atom.cm$^{-2}$), it can bias studies.
 The far-infrared fluxes observed in either the ISOPHOT or the IRAS HiRes maps represent the emission from the SMC dust, but also from this Galactic foreground.
 However, this Galactic component of the emission is not observed in the ATCA/Parkes data, due to a velocity cut and spatial filtering.
 To study the SMC dust properties, it was thus necessary to remove this foreground component (see Sect. \ref{sec:remov_gal}).
 In this step, we used a Galactic \ion{H}{i} column density map constructed from data of the Parkes narrow-band \ion{H}{i} survey of the tidal arms of the Magellanic system \citep{BKS00}. The column density map of the \ion{H}{i} Galactic foreground was obtained by integrating the emission from -60 to +50~km/s, outside the SMC velocities. The resolution is $15\arcmin$. Due to side-lobes effects, the map obtained can slightly overestimate the column density by 10 to 20 \%.

\section{Photometric consistency}\label{sec:phot_cons}

DIRBE provided infrared absolute sky brightness maps in 10 bands in the wavelength range 1.25 to 240~$\mathrm{\mu}$m. This absolute photometry is used to check ISOPHOT mosaic and IRAS fluxes.

\subsection{ISOPHOT data}

\begin{figure}
	\resizebox{\hsize}{!}{\includegraphics{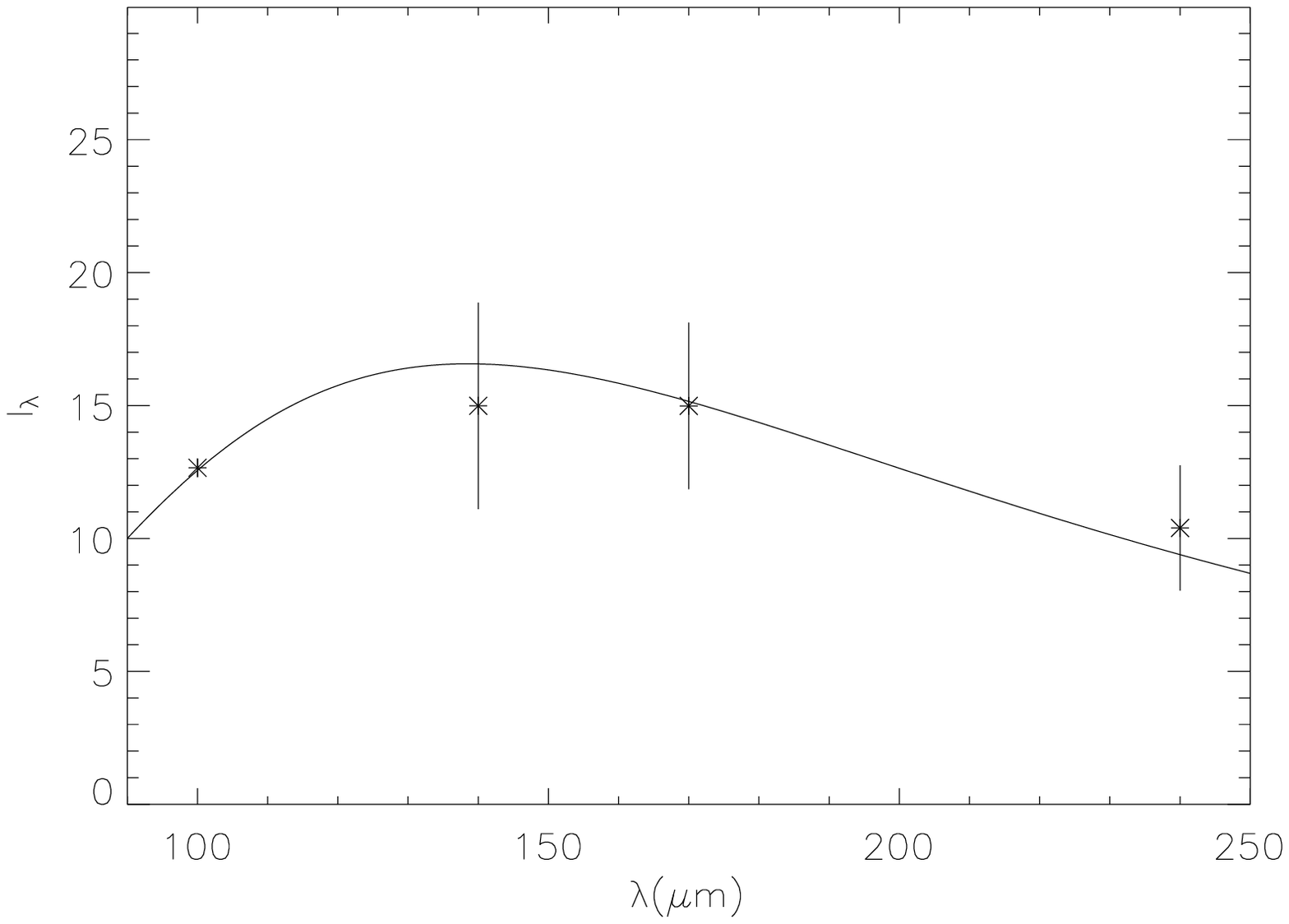}}
	\caption{Comparison of the DIRBE-convolved ISOPHOT fluxes with the DIRBE ones (in MJy/sr), after an integration on the whole SMC. The modified black body traced was fit on the DIRBE values only and correspond to a temperature of 20.6~K and an emissivity of $\epsilon_H(250\mu\mathrm{m})=3\cdot10^{-28}\mathrm{cm}^{2}$ for $10^{20}$ hydrogen atoms.cm$^{-2}$. We see that the ISOPHOT value at 170~$\mathrm{\mu m}$ is consistent with the DIRBE photometry.\label{fig:calib_dirbe}}
\end{figure}

We checked if the ISOPHOT 170~$\mathrm{\mu}$m flux is consistent with the spectral energy distributions obtained from the 100, 140 and 240~$\mathrm{\mu}$m DIRBE fluxes in the SMC. To do this we convolved the ISOPHOT map with the DIRBE beam. This gives 18 ISOPHOT fluxes in the SMC as if they were observed by DIRBE and we compare ISOPHOT and DIRBE fluxes by plotting the spectral energy distribution and fitting it with a modified black body with a spectral index of 2 on the DIRBE values. 
Fig. \ref{fig:calib_dirbe} shows that the DIRBE-convolved ISOPHOT flux integrated over the 18 positions observed by DIRBE, is consistent  within the errors bars with the best modified black-body fit (obtained for a temperature of 20.6K, an emissivity of $\epsilon_H(250\mu\mathrm{m})=3\cdot10^{-28}\mathrm{cm}^{2}$ for $10^{20}$ hydrogen atoms.cm$^{-2}$). Please note that these values correspond to an average over the whole galaxy and were obtained only for calibration purposes.

We have also plotted the individual fit values at $170~\mathrm{\mu m}$ versus the corresponding 18 ISOPHOT values. 
This plot shows a significant scatter but is consistent with a linear correlation with unity slope and zero offset.

A comparison with the IRAS high resolution data was also performed to check possible photometric errors introduced by the interpolation process. The correlation between IRAS and ISOPHOT data does not change when we restrict the comparison to the pixels measured by ISOPHOT rather than the full interpolated map. The interpolation required to fill the holes in the measurements does not seem to bias the image comparisons.

\subsection{IRAS data}\label{sec:phot_iras}

The IRAS HiRes maps have the same photometric calibration as the IRAS ISSA maps.
The IRAS detector had a dwell-time dependent responsivity change: the gain changes as a function of source size. 
This effect is known as the `` AC/DC effect''. 
A first set of corrections were taken into account in the IRAS data products. These corrections were later found to be insufficient. On the basis of a comparison of the IRAS and DIRBE sky brightness for extended emission, \citet{WGC+94} recommended applying additional correction factors of 0.88, 1.01, 0.82 and 0.74 at 12, 25, 60 and 100~$\mathrm{\mu}$m respectively.

However, these correction factors are only valid for spatial scales greater than 40\arcmin.
 This is not true in the SMC where structures on various spatial scales are present.
 \citet{SSV+00} have also compared DIRBE and IRAS data by integrating fluxes over 6.25$\degr^2$ in the SMC.
 They found that integrated flux densities measured by DIRBE for 60 and 100~$\mathrm{\mu}$m are higher than IRAS fluxes by about 10 to 20\%, which is an opposite correction to that observed by \citet{WGC+94}. 

We have compared the IRAS (convolved to the DIRBE resolution) and the DIRBE 100~$\mathrm{\mu}$m maps on a $10\degr \times 10\degr$ scale around the SMC to understand this discrepancy. 
We see that the \citet{WGC+94} correction factor of 0.74 is only truly valid outside the galaxy and increases in the SMC up to a maximum value in the north of the optical bar (N66 region).
For the diffuse medium of the SMC, the \citet{WGC+94} correction factor is sufficient. After correction, we estimate that the 100~$\mathrm{\mu}$m IRAS brightness for the diffuse medium is underestimated by less than 15\%. 

\subsection{Galactic foreground removal}\label{sec:remov_gal}

Unlike the ATCA/Parkes \ion{H}{i} map, the far-infrared emission maps (ISOPHOT and IRAS) trace both the dust emission in the SMC and the dust in our Galaxy. To disentangle these two components in the far-infrared emission, we used a column density map of the \ion{H}{i} Galactic foreground presented in Sect. \ref{sec:data_gal_fil}. 

Dust grains absorb energy from the UV/optical radiation field that they re-emit in the infrared wavelength as thermal emission. 
Large grains are at a fixed temperature and their emitted intensity can be expressed as:
\begin{equation}\label{equ:emi_pouss}
I_\nu = N_H\epsilon_H(\nu)B_{\nu}(T_{dust})
\end{equation}
where $N_H$ is the column density of hydrogen and $\epsilon_H(\nu)$ is the emissivity of grains per hydrogen atom. Assuming that the emissivity of dust grains in the far-infrared follows a power law:

\begin{equation}\label{equ:loi_emis}
\epsilon_H(\nu)=\epsilon_H(\nu_0)(\frac{\nu}{\nu_0})^{\beta}
\end{equation}

and taking $\beta=2$ (as found for Galactic emission \citep{LABP98,LAB+99} for wavelengths between 100~$\mathrm{\mu}$m and 500~$\mathrm{\mu m}$).

Using a temperature of 17.5~K and an emissivity per hydrogen atom of $10^{-25}$ cm$^2$ \citep{BAB+96}\footnote{We checked that the observed IRAS and DIRBE far-infrared ratios around the SMC are consistent with these values.} for the Galactic dust, we computed the Galactic thermal emission and removed this foreground emission from the IRAS and ISOPHOT maps to obtain the emission of the SMC dust only. At 12, 25 and 60~$\mathrm{\mu m}$, we used the $I_{12}/I_{100}$, $I_{25}/I_{100}$ and $I_{60}/I_{100}$ ratios computed from IRAS maps outside the SMC, together with the computed foreground emission map at 100~$\mathrm{\mu m}$ to remove the Galactic foreground emission at these wavelengths. However, this foreground emission remains in the data noise at 12, 25 and 60~$\mathrm{\mu m}$.

\section{Spatial image comparison}\label{sec:projres}

\subsection{Projection and resolution}

The data set presented in Sect. \ref{sec:data} is composed of maps with different fields of view and different resolutions. We projected all of them onto a common grid: that of IRAS HiRes maps ($2\degr\times2\degr$ field of view centered at  $00^h 57^m 29.12^s  -72\degr 32\arcmin 29.2\arcsec$ (J2000) with 10~$\arcsec$ pixel size). These maps were inter-compared by pairs.

To compare the ISOPHOT map with one of the HiRes maps, the projected and re-sampled ISOPHOT map was convolved with a spatial varying kernel in order to be at the HiRes resolution when the point spread function of ISOPHOT was smaller than the one of HiRes \citep{KM01}.  
The HiRes map was also convolved with a spatially varying kernel to match the ISOPHOT resolution when the HiRes beam was smaller than the ISOPHOT one.

To compare the HI column density map with the ISOPHOT map, no convolution was necessary as long as their resolutions were comparable.

\begin{figure*}
	\centering
	\includegraphics[angle=90,width=9cm]{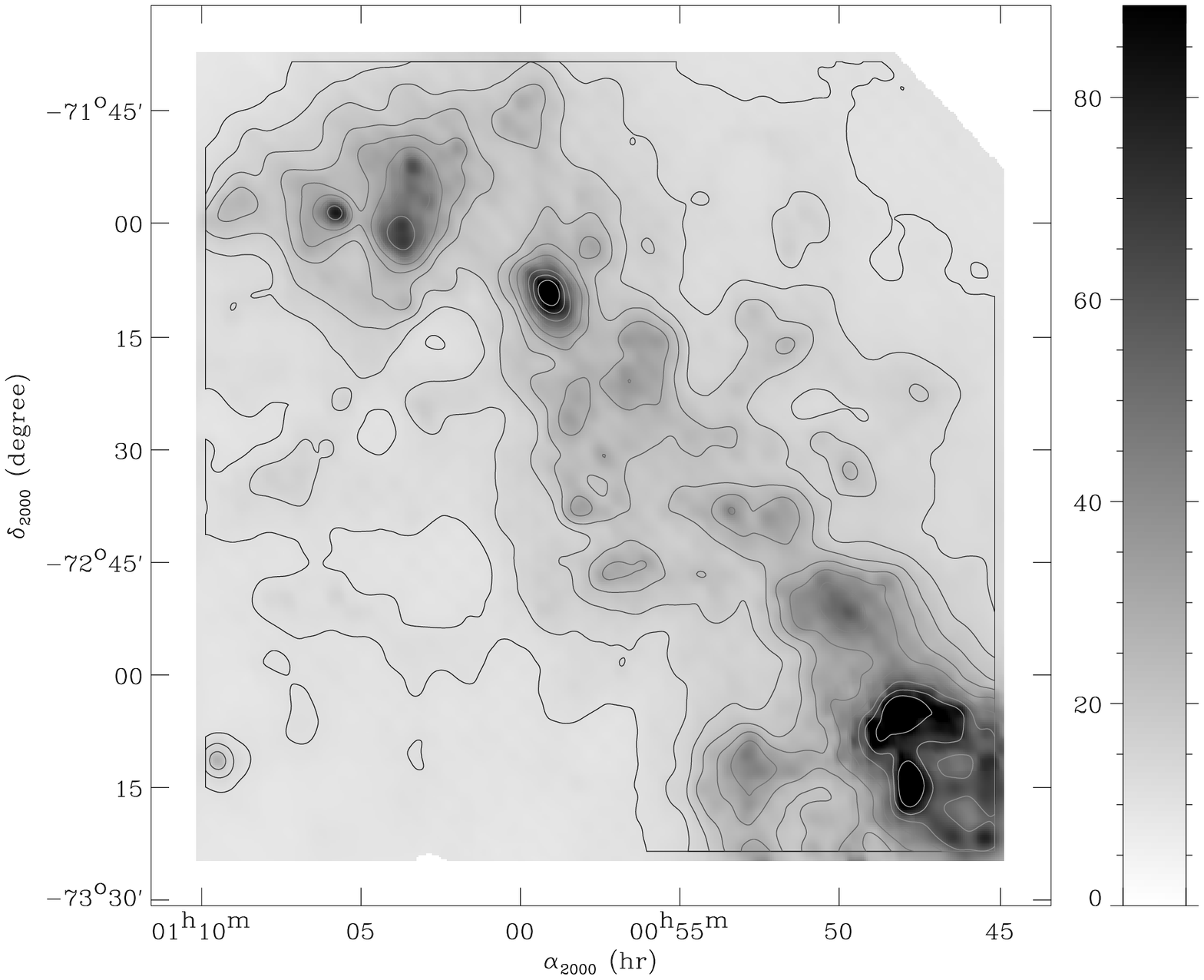}\includegraphics[angle=90,width=9cm]{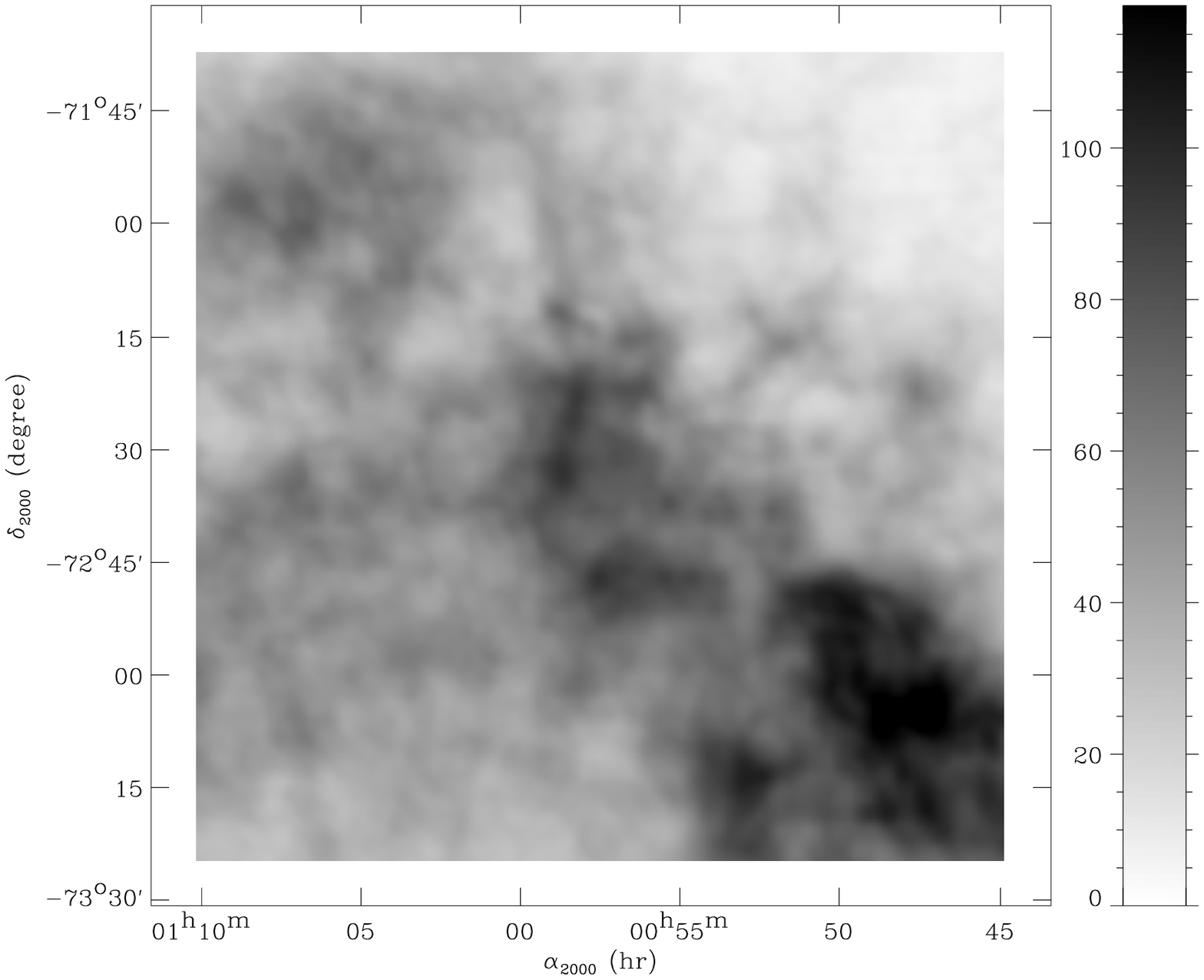}
	\caption{ISOPHOT map (left panel, in MJy/sr) and ATCA/Parkes \ion{H}{i} column density map (right panel, in $10^{20}$ Hydrogen atom/cm$^2$) after projection on the IRAS HiRes grid, and at the same resolutions. For the ISOPHOT map, logarithmic contours are overplotted at fluxes 11.2, 14.1, 17.8, 22.4, 28.2, 35.5, 44.7, 56.2, 70.8 and 89.1 MJy/sr. \label{fig:map_fin}}
\end{figure*}

Fig. \ref{fig:map_fin} shows two of the transformed maps: the ISOPHOT map  degraded to the IRAS 100~$\mathrm{\mu}$m resolution and the projected ATCA/Parkes map.

\subsection{Colors from scatterplots}\label{sec:ratio}

To characterize the diffuse medium (i.e. the extended region where the infrared emission--\ion{H}{i} column density correlation is homogenous) spectral energy distribution, we developed a comparison method that is not biased by point sources and is also not affected by offsets in the data calibration.

We work with two maps (obtained after the treatment presented in Sect. \ref{sec:projres}).
The general correlation between the two maps is first quantified by fitting a straight line to the scatterplot of the pixel-to-pixel comparison. We deduce a first ratio value and an error $\sigma$ given by the dispersion in the correlation, which enables us to filter out localized sources (i.e. star-forming regions) by removing pixels that are 3$\sigma$ above the correlation (there is no points below $-3\sigma$).

From the `` filtered'' maps, we compute a ratio map by linearly fitting the pixel correlation in the scatterplot for a sliding box of 50$\times$50 pixels.
 This corresponds to a 5.5$\times$5.5 beams box which is a minimum size to have independent data points.
 The size of the box could not be greater, or we would not trace the local variations.
In some cases there are too few points inside a box (too many points above 3$\sigma$) to do a meaningful fit. We assign no value to these positions. 
We also skip positions where the correlation coefficient is too low.
 Adjusting a gaussian on the histogram of the ratio values gives us a reference ratio and the associated dispersion.
This dispersion contains photometric and fitting errors but also a true scatter in the ratio. Both phenomena cannot be easily disentangled.

 The spatial distribution and the histogram with the gaussian fit obtained for the $I_{100\mathrm{\mu m}}/I_{170\mathrm{\mu m}}$ ratio are presented in Fig. \ref{fig:hist_100s170} as an example.
We see that the comparison method avoids large areas in the main body (bar) of the SMC. These regions correspond to highly structured star-forming sites. In these regions, either the filtering technique has left too few independent positions in one box to compute a significant correlation or no clear correlation is observed.

\section{Results}\label{sec:Results}

\subsection{Infrared colors}\label{sec:IRcolors}

The method presented in Sect. \ref{sec:ratio} was applied to the IRAS HiRes and ISOPHOT maps to compute the $I_{100\mathrm{\mu m}}/I_{170\mathrm{\mu m}}$, $I_{60\mathrm{\mu m}}/I_{170\mathrm{\mu m}}$, $I_{25\mathrm{\mu m}}/I_{170\mathrm{\mu m}}$, $I_{12\mathrm{\mu m}}/I_{170\mathrm{\mu m}}$ ratios. Adjusting a gaussian on the histograms gives us reference ratios and their associated dispersions:
\begin{equation}
\frac{I_{100\mathrm{\mu m}}}{I_{170\mathrm{\mu m}}}=1.0\pm0.2
\label{equ:color1}
\end{equation}
\begin{equation}
\frac{I_{60\mathrm{\mu m}}}{I_{170\mathrm{\mu m}}}=0.5\pm0.1
\end{equation}
\begin{equation}
\frac{I_{25\mathrm{\mu m}}}{I_{170\mathrm{\mu m}}}=0.02\pm0.01
\end{equation}
\begin{equation}
\frac{I_{12\mathrm{\mu m}}}{I_{170\mathrm{\mu m}}}=0.011\pm0.006
\end{equation}

\begin{figure*}
	\includegraphics[width=9cm]{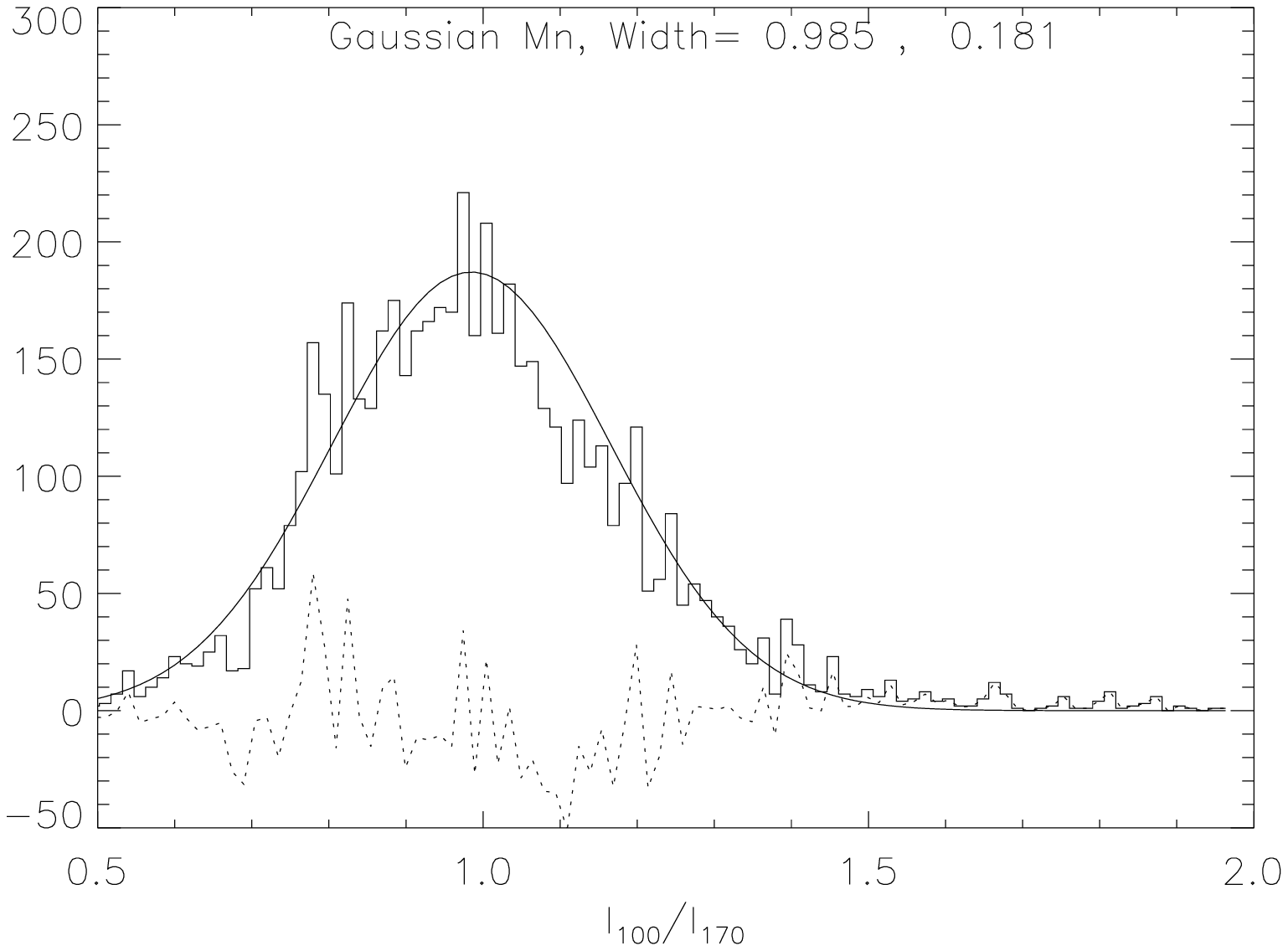}\includegraphics[height=9cm,angle=90]{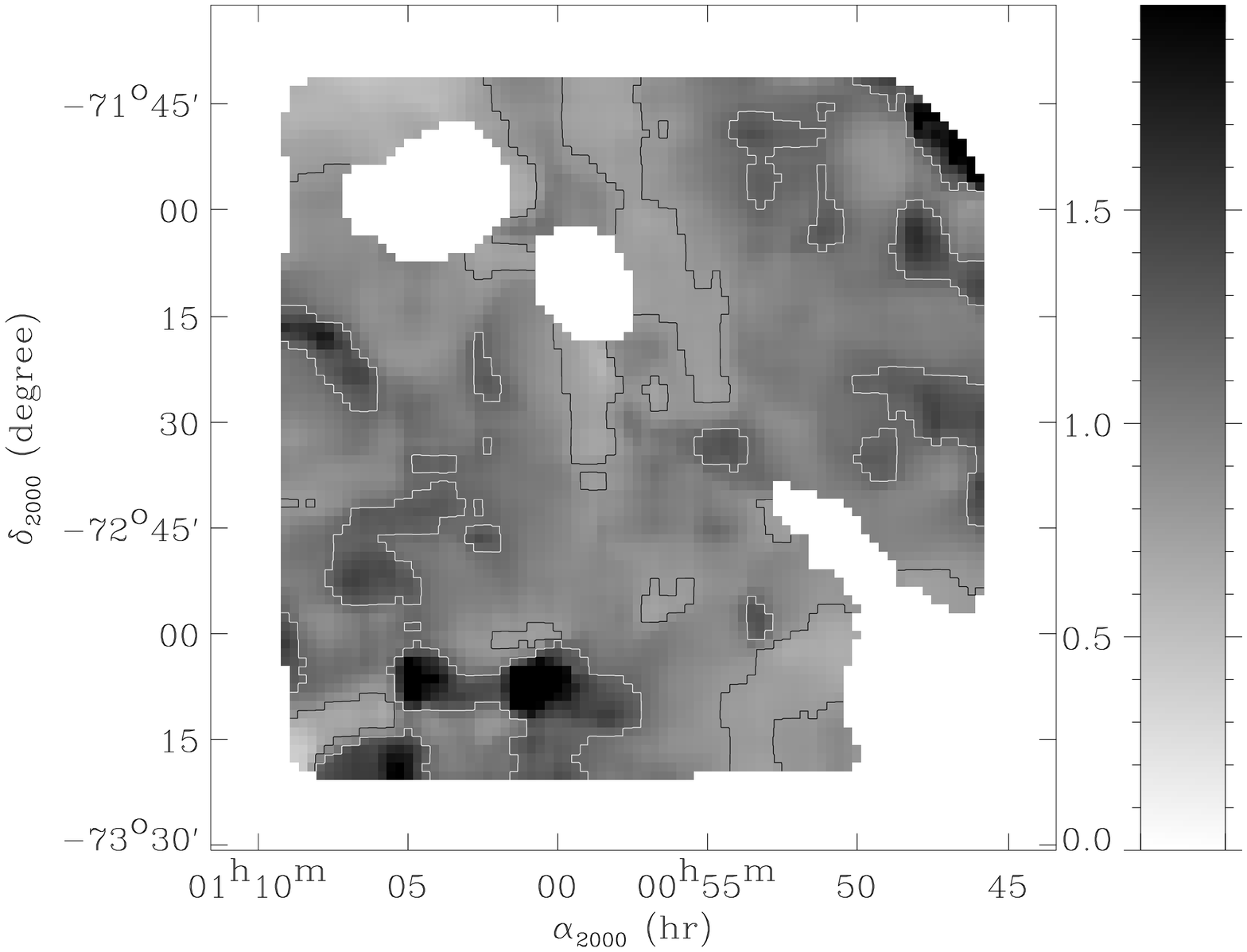}
	\caption{Histogram and spatial distribution of the 100--170~$\mathrm{\mu m}$ colors. For the histogram, the gaussian fit and the residuals are presented as continuous and dotted lines respectively. For the map of the 100--170~$\mathrm{\mu m}$ colors, contours correspond to the 0.8 (black) and 1.2 (white) levels (one sigma limits of the gaussian fit). It should be noticed that the noise in the color map is not homogeneous.\label{fig:hist_100s170}}
\end{figure*}

Using equation \ref{equ:emi_pouss} and \ref{equ:loi_emis} we can derive from the $\frac{I_{100~\mathrm{\mu m}}}{I_{170~\mathrm{\mu m}}}$ ratio map the big grain equilibrium temperature distribution in the diffuse medium in the Small Magellanic Cloud.
This leads to a reference temperature of: $T_\mathrm{dust}=22\pm2~\mathrm{K}$.
 Note that this temperature is different from the one that would be inferred from the ratios of the mean $I_{100~\mathrm{\mu m}}$ and $I_{170~\mathrm{\mu m}}$ fluxes. 
\citet{WKL+04} found a mean temperature for the main body of the SMC of $20.9\pm 1.9\mathrm{K}$, which is lower than the temperature we find for the diffuse medium. Their mean value falls within our range of temperatures, which is expected because the regions we study overlap.
The difference in mean values could be due to the presence of cold atomic and molecular clouds in the main body of the SMC.
Our studies seem to be consistent with each other and are more complementary than comparable.
 
The large dust grain temperature is related to the radiation field intensity and will be discussed in Sect. \ref{sec:discuss}.

\subsection{Gas to dust correlation}\label{sec:gas-dust-col}

\begin{figure}
	\resizebox{\hsize}{!}{\includegraphics[width=0.5\textwidth]{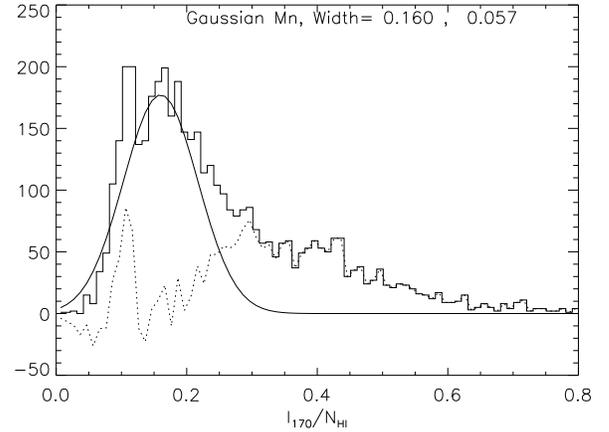}}
	\caption{Histogram of the 170 micron flux to \ion{H}{i} column density ratio (in MJy.cm$^2$/sr). The gaussian fit and the residuals are presented as continuous and dotted lines respectively\label{fig:histo_170shi}}
\end{figure}

From the ISOPHOT and the ATCA/Parkes maps, using the method described Sect. \ref{sec:ratio}, we found the distribution presented on Fig. \ref{fig:histo_170shi}. This distribution exhibits a main peak and a distribution tail. The peak is characterized by:

\begin{equation}
\frac{I_{170\mathrm{\mu}\mathrm{m}}}{N_{HI}}=0.17\cdot10^{-20}\pm0.06\cdot10^{-20} MJy.sr^{-1}.cm^2
\label{equ:irshi}
\end{equation}

\begin{figure}
	\resizebox{\hsize}{!}{\includegraphics[angle=90,width=0.4\textwidth]{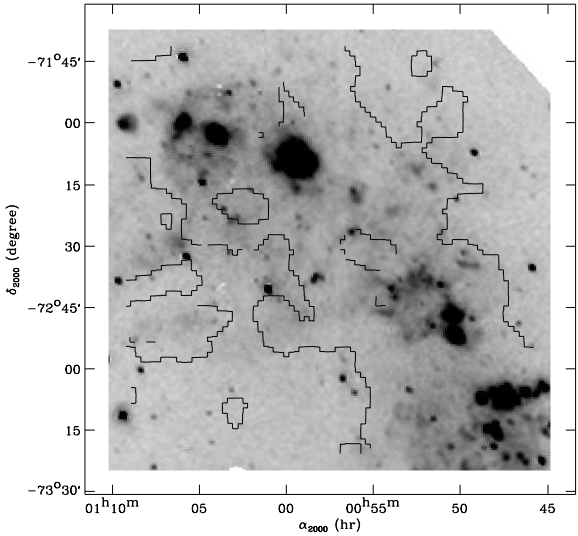}}
	\caption{Uncalibrated \ion{H}{$\alpha$} map. As an illustration, the contour line corresponds to $I_{170~\mathrm{\mu m}}/N_{HI}=0.23$ MJy.cm$^2$/sr (see equation \ref{equ:irshi}) as in figure \ref{fig:map_170shi}. \label{fig:ha_limrap}}
\end{figure}

Since the regions used to define the ratios are not exactly the same, the determination of the characteristics of the diffuse medium could be biased. We checked that this is not the case. If we use the region where $I_{170\mathrm{\mu}\mathrm{m}}/N_{HI}$ is computed to determine the colors, we find that the derived quantities are not affected by more than $15\%$. 
We see that the characteristics of the diffuse medium are not biased by the definition of the limits of the studied region.
Comparing the region defined by equation \ref{equ:irshi} with the \ion{H}{$\alpha$} distribution, we see in Fig. \ref{fig:ha_limrap} that the regions corresponding to the tail in the $I_{170\mathrm{\mu}\mathrm{m}}/N_{HI}$ (located mainly in the main body) are regions where the diffuse \ion{H}{$\alpha$} emission is important.
Comparing the map in Fig. \ref{fig:hist_100s170}  and Fig. \ref{fig:ha_limrap}, we also see that the regions where no ratios can be computed correspond to important star-forming sites. 

Assuming that all the dust is in the \ion{H}{i}, we use the $I_{170\mathrm{\mu m}}/N_{HI}$ ratio map and the temperature map computed above to obtain an histogram of the emissivity per hydrogen atom of the dust grains in the diffuse medium. We find a reference emissivity at 170~$\mathrm{\mu}$m of:

\begin{equation}
\epsilon_H(\lambda)=7.3\cdot10^{-27}\pm3.2\cdot10^{-27}\mathrm{cm}^{2}
\end{equation}

The error bar takes into account uncertainties in the temperature and $I_{170~\mathrm{\mu m}}/N_{HI}$.
If the emissivity follows a power law with a spectral index of 2, the emissivity per \ion{H}{i} atom in the SMC at 250~$\mathrm{\mu}$m is $\epsilon_H(\lambda)=3.4\cdot10^{-27}\pm 1.5\cdot10^{-27}(\lambda/250~\mathrm{\mu m})^{-2}\mathrm{cm}^{2}$. This corresponds to $\kappa_\nu=0.21(\lambda/250~\mathrm{\mu m})^{-2} \mathrm{cm}^2\mathrm{g}^{-1}$. Assuming similar dust optical properties in the SMC and the Galaxy, the emissivity is related to the GDR and the ratio between hydrogen column density and $\mathrm{E(B-V)}$ color excess by:

\begin{equation}
\frac{\epsilon_H(SMC)}{\epsilon_H(Galaxy)}=\frac{\frac{M_{dust}}{M_{gas}}(SMC)}{\frac{M_{dust}}{M_{gas}}(Galaxy)}=\frac{\frac{N(H_I)+N(H_2)}{E(B-V)}(Galaxy)}{\frac{N(H_I)+N(H_2)}{E(B-V)}(SMC)}
\end{equation}

The solar neighborhood value for the emissivity is $10^{-25}$~cm$^2$ at 250~$\mathrm{\mu}$m \citep{BAB+96}. The inferred SMC GDR is thus 30 times greater than in the Galaxy. 
\citet{SSV+00} have presented a similar study using the same \ion{H}{i} data together with HIRAS data. They came to the same conclusion that the GDR is a factor 30 higher in the SMC than in our Galaxy. However, this agreement might be a coincidence since 1) we did not use the same calibration of the IRAS data, 2) their temperature estimation based on the $60/100\mathrm{\mu m}$ color is higher than ours, 3) the GDR they present is averaged over the whole galaxy (including star forming regions) and 4) their GDR is compared to a Galactic GDR of 375 instead of the standard value of 100\footnote{This value can be checked with the interstellar abundance standards taken from table 2 of \citet{SM01a} and \citet{SM01b}.}. Using the standard Galactic GDR value, the average SMC GDR found by \citet{SSV+00} becomes 100 times higher than in the Galaxy.
 
Our estimate of the SMC's GDR 30 times higher than in the Galaxy, is significantly larger than the difference in metallicity, which is less than 10 \citep{Duf84}. It is twice the observed difference in the $\frac{N(H_I)+N(H_2)}{E(B-V)}$ values presented by \citet{TSR+02}.

\subsection{Infrared excesses}

\begin{figure}
	\resizebox{\hsize}{!}{\includegraphics[angle=90,width=0.4\textwidth]{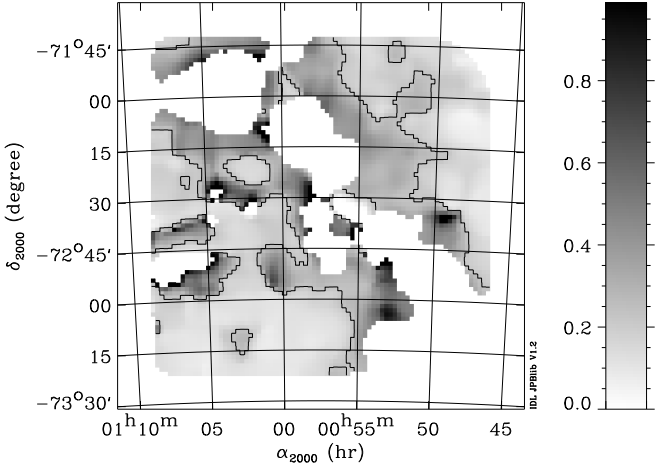}}
	\caption{Spatial distribution of the $I_{170~\mathrm{\mu m}}/N_{HI}$ ratio (in MJy.cm$^2$/sr). Regions where no ratio is computed (see Sect. \ref{sec:ratio}) are left blank. The contour line corresponds to a  $I_{170~\mathrm{\mu m}}/N_{HI}=0.23$ MJy.cm$^2$/sr level. We see that the IR excesses are concentrated near the optical bar.\label{fig:map_170shi}}
\end{figure}

A noticeable result of the ISOPHOT and \ion{H}{i} comparison is the presence of a tail in the 170~$\mathrm{\mu}$m/\ion{H}{i} histogram (see Fig. \ref{fig:histo_170shi}).
This tail corresponds to pixels in the SMC optical bar with excess 170~$\mathrm{\mu m}$ emission with respect to $N_{HI}$ (see Fig. \ref{fig:map_170shi}).
We checked that the mean temperature for this region of excess is the same as that for the diffuse medium. 
 These excesses may be due either to a variation in the dust abundances between the diffuse medium and the bar, or due to the presence of cold \ion{H}{i} or H$_{2}$ gas.
We believe that the excesses are real but it is difficult to infer an amount of cold \ion{H}{i} and molecular gas from it.
 Firstly, it is difficult to disentangle these two possible contributions. 
Secondly, where some of these excesses are due to dust in cold \ion{H}{i} or H$_{2}$ we do not expect them to be correlated with \ion{H}{i} emission and another method is needed to quantify them. It is thus beyond the scope of this paper to translate these excesses into an estimate of cold gas that could be compared to other studies where this is attempted \citep{SSV+00}.
Finally, we feel that the photometric uncertainties in the ISOPHOT and IRAS data are an additional difficulty, especially near star-forming regions.
 Further studies with the Spitzer satellite could address this question with better photometry and angular resolution.

\subsection{SMC dust properties in diffuse \ion{H}{i}: modeling the spectral energy distribution}\label{sec:diffhi}

The different intensity ratios obtained in Sect. \ref{sec:IRcolors} enable us to trace the spectral energy distribution of dust in the diffuse medium of the SMC and to compare it with that of the solar neighborhood.
The difference in temperature and in abundances is seen Fig. \ref{fig:spectrum} as a shift along the wavelength and brightness axis respectively.
 The error bars include true variations in the dust properties across the SMC diffuse interstellar medium which are correlated over wavelengths.

\begin{figure}
	\resizebox{\hsize}{!}{\includegraphics{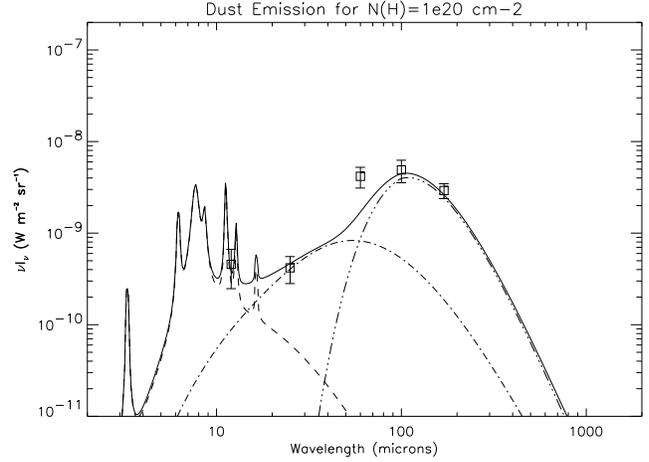}}
	\caption{Best fit for the dust infrared emission spectrum, obtained with the Desert et al. model. The brightness corresponds to an interstellar gas column density of $10^{20}$cm$^{-2}$. The IRAS and ISOPHOT data are represented by squares. The fit obtained with the model (continuous line) corresponds to the emission of three components: PAHs(dashed line), very small grains (mixed line) and big grains (dot-dot-dot-dashed line).\label{fig:spec_mod}}
\end{figure}

\begin{figure*}
	\resizebox{\hsize}{!}{\includegraphics{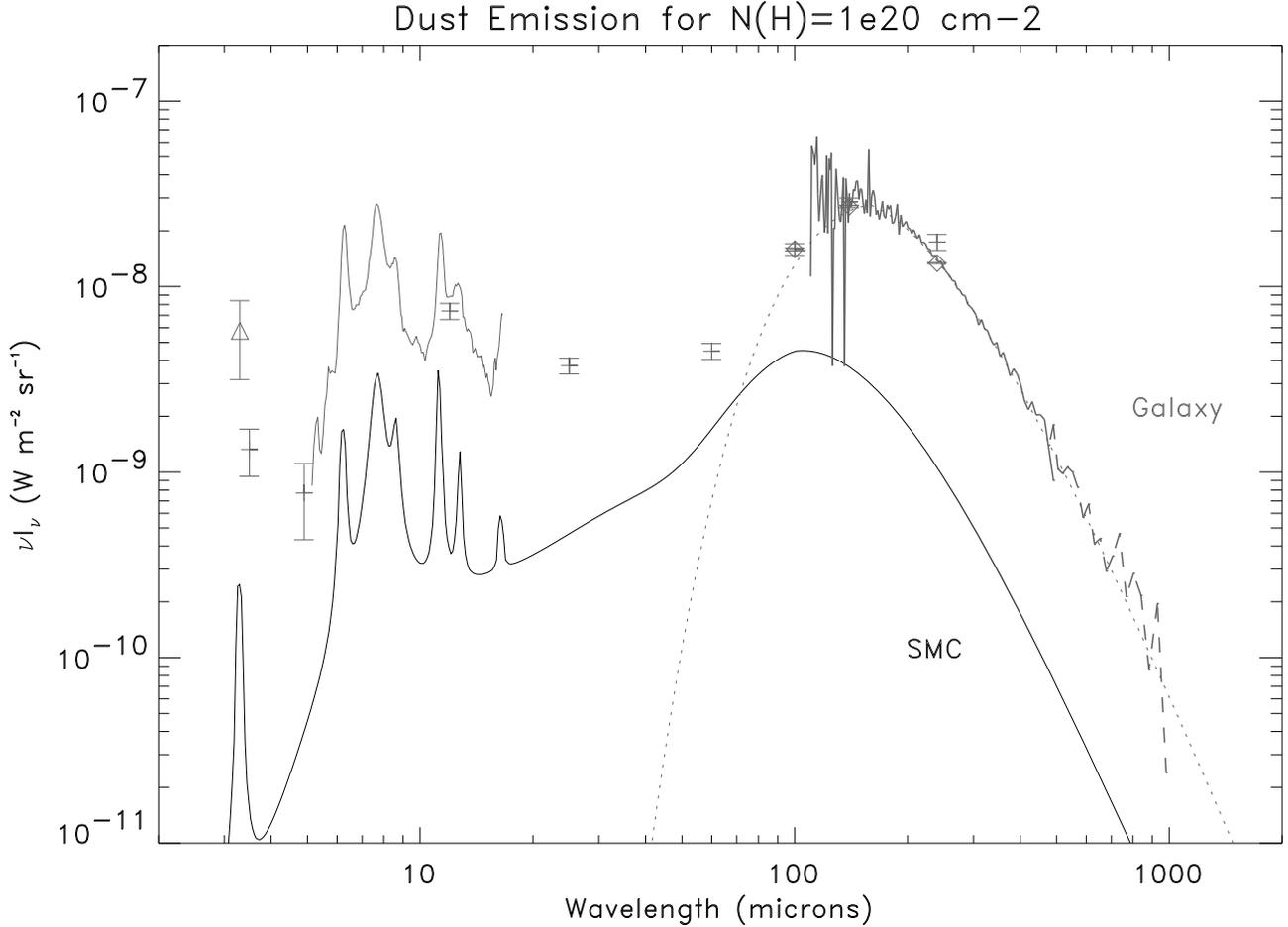}}
	\caption{Infrared spectral energy distribution of dust in the diffuse medium for the SMC and the Solar Neighborhood. The brightness corresponds to an interstellar gas column density of $10^{20}$cm$^{-2}$. The Galactic spectrum (in grey) is composed in the far-IR by COBE/FIRAS data from \citet{BAB+96} (with the corresponding spectrum from grains in thermal equilibrium indicated by the dotted line), and at shorter wavelengths  by the COBE/DIRBE \citep{BBD+94} and Arome balloon data (triangle at 3.3~$\mu$m) \citep{GLP+94}. An ISOCAM spectrum \citep{BRA+96} scaled to the 12~$\mu$m brightness is also shown. The SMC spectrum (in black) is the best fit obtained with the Desert et al. model, same as in figure \ref{fig:spec_mod}. \label{fig:spectrum}}
\end{figure*}

We used the dust model of Desert et al. (1990) to quantify the grain dust abundances. 
In order to model the spectral energy distribution (SED), we need to estimate the interstellar radiation field (ISRF) in the diffuse medium of the SMC.
In Sect. \ref{sec:IRcolors} we found the big dust grain equilibrium temperature to be 22~K. Using the dust thermal equilibrium temperature in the solar neighborhood of 17.5~K \citep{BAB+96} as a reference, the dust temperature in the SMC leads to a radiation field $\chi_{SMC}=T_{dust}^{4+\beta}(SMC)/T_{dust}^{4+\beta}(Galaxy)=4$.
This value is consistent with the estimate of the ISRF made from massive star counts by \citet{Leq79}. \citet{VLMR80} also found the same ratio from UV measurements.
 We used the \citet{MMP83} 10~kpc Galactic ISRF and scaled it by a factor $\chi=4$ for the modeling.
We have adjusted only the abundances of PAHs, very small grains (VSG) and big grains. For the other parameters defining the size distribution, we have kept the Galactic values.
The best fit is presented on Fig. \ref{fig:spec_mod}, it corresponds to under-anbundant PAHs, very small grains (VSG) and big grains (BG) compared to the Galaxy (36, 23 and 30 times less respectively).
Mass abundances relative to hydrogen, compared with Galactic values (Desert et al. 1990) are summarized in table \ref{tab:model}. 

\begin{table}[!htbp]
	\caption{Mass abundances relative to hydrogen in the SMC and the Galaxy. The grain size distribution is assumed to be the same in both cases.\label{tab:model}}
	\begin{center}
	\begin{tabular}{l l l }
		\hline
		Component&SMC &Galaxy\\
		&$\frac{m}{m_H}$&$\frac{m}{m_H}$\\
		\hline
		PAH &$1.2\cdot 10^{-5}$ &$4.3\cdot 10^{-4}$\\
		VSG &$2.0\cdot 10^{-5}$ &$4.7\cdot 10^{-4}$\\
		BG &$2.2\cdot 10^{-4}$ &$6.4\cdot 10^{-3}$\\
		\hline
	\end{tabular}
	\end{center}
	
\end{table}

The modeling confirms the GDR estimated in Sect \ref{sec:gas-dust-col} to be around 30~times greater than in the Galaxy.
Within the uncertainties, PAHs and very small grain abundances are also consistent with this factor. 

The model is clearly below the 60~$\mathrm{\mu m}$ point.
This could be due to our crude estimate of the ISRF. 
We have used a single ISRF while our data analysis shows temperature variations across the SMC diffuse medium.
 To take the variations in the radiation field into account, we transformed the temperature distribution obtained in Sect. \ref{sec:IRcolors} into a $\chi_{SMC}$ distribution.
 For each value of the distribution, we computed a spectral energy distribution (SED) of the dust emission with the \citet{DBP90} model, using the Mathis 10~kpc ISRF multiplied by the corresponding $\chi$ as an input ($\chi$ is a constant for the whole spectrum).
 We then computed the resulting total SED by adding the individual SEDs with weights according to the $\chi$ distribution shape.
The resulting fit is not sufficiently broadened to encompass the 60~$\mathrm{\mu m}$ point. In addition, the mass abundances of dust grains have to be reduced by 8, 5 and 7 for the PAH, VSG and BG components respectively.
Due to the high non-linearity of the Planck curve, even a small broadening of the $\chi$ distribution tends to reduce the abundances of dust grains significantly. These values are thus indicatives only but they show that the mass abundances represented in table \ref{tab:model} are more likely to be upper limits and that the emissivity could be even lower.

\begin{figure}
	\resizebox{\hsize}{!}{\includegraphics[width=0.4\textwidth]{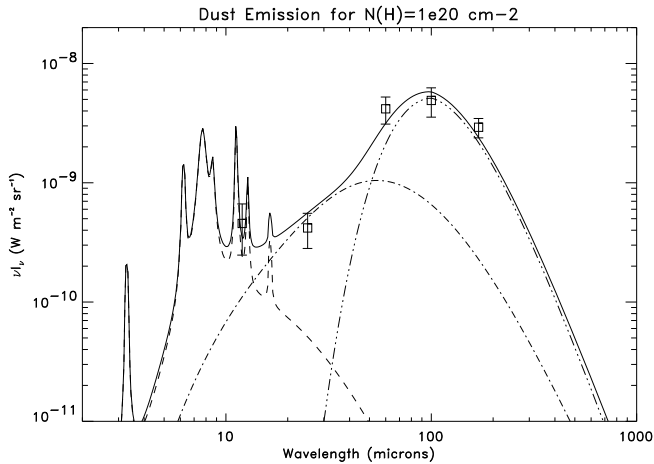}}
	\caption{Dust infrared emission spectrum fit obtained with the Desert et al. model to adjust the $60\mu\mathrm{m}$ excess. The IRAS and ISOPHOT data are represented by squares. This spectrum is obtained with a single ISRF scaled by $\chi=10$ and reduced dust grain mass abundances for PAHs (dashed line), very small grains (mixed line) and big grains (dot-dot-dot-dashed line).\label{fig:model_chi10}}
\end{figure}

It is possible to fit the $60~\mathrm{\mu m}$ value with a single ISRF scaled by $\chi=10$ (see Fig. \ref{fig:model_chi10}).
 In this case, the dust grain mass abundances must then be reduced by 3, 2 and 2 for the PAHs, very small grains and big grains respectively.
 Such a high ISRF corresponds to a dust temperature of 25.6~K, which is higher than the estimate obtained in Sect.~\ref{sec:IRcolors} but still compatible with the $100~\mathrm{\mu m}$ and the $170~\mathrm{\mu m}$ fluxes.
 This higher temperature would lead to an emissivity of $2.4\cdot10^{-27} \mathrm{cm}^2$ at $250~\mathrm{\mu m}$, which is even lower than the estimation obtained in Sect. \ref{sec:gas-dust-col} and results in a GDR 40 times higher than in the Galaxy. A higher radiation field could thus account for the $60\mu\mathrm{m}$ excess without contradicting the main conclusion of this study, namely the fact that the GDR in the diffuse SMC ISM is much lower than the solar neighbourhood value scaled by the difference in metallicity.

The observed $60~\mathrm{\mu m}$ excess could also be due to a change in the grain size distribution \citep{GMJ+03}. 
We assumed the Galactic grain sizes distribution to apply in the SMC. 
However, by changing the relative mass contributions of the different grain components, we enlarge the discontinuities in the size distribution between the three components of the model.
It is necessary to further increase these discontinuities to account for the  $60~\mathrm{\mu m}$ excess.

 Further investigations are needed to characterize the SMC ISRF and especially the $60~\mathrm{\mu m}$ excess. The intensity and spectral shape of the ISRF are likely to be different than the Galactic one.

\section{Discussion}\label{sec:discuss}

Studying the gas to dust correlation in the diffuse medium of the SMC (see Sect. \ref{sec:gas-dust-col}), we found a GDR in mass around 30 times greater than in the solar neighborhood, larger by a factor of 3 than the difference in metallicity. This result confirms the value found by \citet{SSV+00} in a similar study.
This lower depletion of heavy elements in dust was independently reported by \citet{WLB+01}, who found Mg and Si to be essentially undepleted in the SMC.
It was also reported for a sample of dwarf irregulars by \citet{LF98}. Our result in the SMC fits with the relation that they observed between GDR and metallicity.

 An interpretation of this relation has been proposed by \citet{HTK02}. With a chemical evolution model, they illustrated the dependence of the GDR on the star formation history. Stars generate dust but also contribute to their destruction by supernova shocks.
 They propose a scenario in which an intermittent star formation history leads to a time variable GDR. At some epochs of the galaxy evolution, the formation of dust grains is less efficient than the destruction process by supernova shocks, so that the GDR no longer scales with the metallicity.
This model could be applied to the SMC with its specific star formation history \citep{HZ03}. It is also necessary to account for a possible difference in the GDR between the diffuse and the denser medium.

Extinction studies lead to significantly smaller value of the GDR, closer to that expected from the Galactic value for a linear dependence with metallicity. \cite{BLM+85} used photometric observations of O-B stars in the SMC to derive their extinction curves. They found a GDR value that is about 8 times the Galactic value, which is consistent with the difference in metallicity. 
More recently, \citet{TSR+02} measured $\frac{N(H_I)+N(H_2)}{E(B-V)}$ for a sample of stars observed with FUSE and found a mean value 16 times lower than the solar Neighborhood value. This is closer but still smaller by a factor 2 than our estimate. The difference may be due to the fact that extinction
studies are based on line-of-sights with high hydrogen column density where dense interstellar medium 
is likely to be present. Most of the stars are in the optical bar and/or 
in star formation regions associated with molecular clouds. 
The difference in the GDRs inferred in our work and extinction studies might thus point to
a dependence on the environment: a decrease of the ratio from the diffuse ISM to  
star forming regions that might arise from the condensation of heavy elements on 
dust grains in dense gas and their destruction by supernova shocks in the diffuse ISM. 
\citet{SSV+00} have also noted this possibility for the SMC, but it still requires confirmation by further studies.

\section{Summary and conclusions}\label{sec:ccl}

This study was made feasible by the availability of 10 ISOPHOT observations at $170~\mathrm{\mu m}$, covering most of the SMC, in the ISO Data Archive. After reduction processes, mapping and Galactic foreground removal, the ISOPHOT data were compared with IRAS maps and a ATCA/Parkes combined \ion{H}{i} column density map to assess the properties of the dust in the diffuse medium of this low metallicity galaxy.

For the dust in the diffuse medium of the SMC, we found a reference equilibrium temperature of $22 \pm 2$~K  and a reference dust emissivity of $3.4\cdot 10^{-27} \pm 1.5\cdot 10^{-27} \big(\frac{\lambda}{250~\mathrm{\mu m}}\big) \mathrm{cm^2}$ per hydrogen atom. This leads to a gas-to dust ratio 30 times greater than in the Galaxy. This high value reflects the SMC low metallicity, but also requires a lower depletion of dust elements than in the Galaxy. This shows that a simple linear dependence between the GDR and the metallicity does not apply for the diffuse medium of the SMC.

The spectral energy distribution is modeled with the \citet{DBP90} model. The best fit obtained shows a similar decrease for PAHs and very small grains as for big grains. However, a $60~\mathrm{\mu m}$ excess remains that can not be accounted for with simple assumptions. Further constraints, in particular on the ISRF, are needed to understand it.

Our GDR estimate for the diffuse medium of the SMC is a factor 2 to 3 higher than that derived from extinction studies towards bright UV stars with high foreground hydrogen column densities. This difference  supports the notion of variations in the GDR, from diffuse regions where grains are more frequently destroyed by supernovae, to denser and more quiescent regions where grains re-accrete heavy elements. 

\begin{acknowledgements}
We acknowledge S. Stanimirovi\'c for providing us with the ATCA/Parkes combined \ion{H}{i} data and her help in using it, and C. Br\"uns for providing us the Galactic foreground \ion{H}{i} map. This study has been supported by the French national program PCMI (Physique et Chimie du Milieu Interstellaire, CNRS)
\end{acknowledgements}

\bibliographystyle{aa}
\bibliography{./aamnem99,./biblio}

\end{document}